# XRecursive: An Efficient Method to Store and Query XML Documents

Mohammed Adam Ibrahim Fakharaldien, Jasni Mohamed Zain, Norrozila Sulaiman

Faculty of Computer System and Software Engineering, University Malaysia Pahang, Kuantan Malaysia.

**Abstract:** Storing XML documents in a relational database is a promising solution because relational databases are mature and scale very well and they have the advantages that in a relational database XML data and structured data can coexist making it possible to build application that involve both kinds of data with little extra effort . In this paper, we propose an algorithm schema named XRecursive that translates XML documents to relational database according to the proposed storing structure. The steps and algorithm are given in details to describe how to use the storing structure to storage and query XML documents in relational database. Then we report our experimental results on a real database to show the performance of our method in some features.

**Key   words:** *XML, Relational Database, XRecursive,SQl.*

## INTRODUCTION

Today's data exchange between organizations has become challenging because of the difference in data format and semantics of the meta-data which used to describe the data. Now day' XML emerged as a major standard for representing data on the World Wide Web while the dominant storage mechanism for structured data is the relational databases, which has been an efficient tool for storing, searching, retrieving data from different collection of data. The ability to map XML data in relational databases is difficult mission and challenging in the world of all IT organization so there is a need to develop an interfaces and tools for mapping and storing XML data in relational databases.

*XML:*

The extensible Markup Language (XML) is quickly becoming the de facto standard for data exchange over the Internet (Hasan Zafari, Keramat Hasami, M. Ebrahim Shiri, 2010) and now it plays a central role in data management, transformation, and exchange. Since it s introduction to industry in the Late 1990s, XML (M.A. Ibrahim Fakharaldien, Jasni Mohamed Zain, and Norrozila Sulaiman 2011) has achieved widespread support and adoption among all the leading software tools, server, and database vendor s. As importantly, XML has become the lingua franca for data by lowering the cost of processing, searching, exchanging, and re-using information. XML provides a standardized, self-describing means for expressing information in a way that is readable by humans and easily verified, transformed, and published, the hot topic is to seek the best way for storing XML documents in order to get high query processing efficiency(Liu Sainan, Liu Caifeng, Guan Liming,2009). In addition, data can be transmitted to remote services anywhere on the Internet using XML-based Web services to take advantage of the new ubiquity of connected software applications. The openness of XML (Augeri, C. J., Bulutoglu, D. A., Mullins, B. E., Baldwin, R. O., and Baird, L. C.2007) allows it to be exchanged between virtually any hardware, software, or operating system. Simply put, XML opens the door for information interchange without restriction.

*Relational Databases:*

Today, the dominant storage mechanism for structured enterprise data is the relational database, which has proven itself an efficient tool for storing, searching for, and retrieving information from massive collections of data. Relational databases specialize in relating individual data records grouped by type in tables(M.A. Ibrahim Fakharaldien, Siti Normazial Ihsan, Jasni Mohamed Zain, 2010). Developers can join records together as needed using SQL (Structured Query Language) and present one or more records to end-users as meaningful information. The relational database model revolutionized enterprise data storage with its simplicity, efficiency, and cost effectiveness. Relational databases have been prevalent in large corporations since the 1980s, and they will likely remain the dominant storage mechanism for enterprise data in the foreseeable future. Despite these strengths, relational databases lack the flexibility to seamlessly integrate with other systems, since this was not historically a requirement of the database model (Reed, D. 2008). In addition, although relational databases share many similarities, there are enough differences between the major commercial implementations to make

**Corresponding Author:** Mohammed Adam Ibrahim Fakharaldien, Faculty of Computer System and Software Engineering, University Malaysia Pahang, Kuantan Malaysia. E-mail: mohammedfakeraldeen@yahoo.com

developing applications to integrate multiple products difficult. Among the challenges are differences in data types, varying levels of conformance to the SQL standard, proprietary extensions to SQL, and so on.

*Problem Definition:*

For the storage of XML document, the key issue is transforming the tree structure of an XML document into tuples in relational tables (Liew Yue, Jiadong Ren, Ying Qian, 2008). Nowadays, there are more and more data presented as XML document, the need of storing them persistently in a database has increased rapidly while the native–XML databases usually have limited support for relational databases. In recent years, with the popularity of relational databases (RDB), approaches based on RDB to store and manipulate XML data as relational tables but still there is need to manage XML data and relational data seamlessly with similar storage and retrieval efficiencies simultaneously. XML and Relational databases cannot be kept separately because XML is becoming the universal standard data format for the representation and exchanging the information whereas most existing data lies in RDBMS and their power of data capabilities cannot be degraded so the solution to this problem a new efficient methods for storing XML documents in relational database is required.. A new efficient method for storing XML document in relational database is proposed in this paper to face these problems.

*Related Work:*

There are basically three alternatives for storing XML data: in semi-structured databases (R. Goldman, J. McHugh, and J. Widom, 1999) in object-oriented databases (F. Bancihon, G. Barbedette, V. Benzaken, 1988) and in relational systems (I. Tatarinov, S. Viglas, K. Beyer, 2002). Among these approaches, the relational storage approach has attracted considerable interest with a view to leveraging their powerful and reliable data management services. In order to store an XML document in a relational database, the tree structure of the XML document must first be mapped into an equivalent, flat, and relational schema XML documents are then shredded and loaded into the mapped tables. Finally, at runtime, XML queries are translated into SQL, submitted to the RDBMS, and the results are then translated into XML. There is a rich literature addressing the issue of managing XML documents in relational back-ends (M. Ramanath, J. Freire, J. Haritsa, P. Roy, 2003). These approaches can be classified into two major categories as follows:

Schema-conscious approach: This method first creates a relational schema based on the DTD/schema of the XML documents. First, the cardinality of the relationships between the nodes of the XML document is established. Based on this information a relational schema is created. The structural information of XML data is modeled by using primary-key foreign-key joins in relational databases to model the parent–child relationships in the XML tree. Examples of such approaches are Shared-Inlining (J. Shanmugasundaram, K. Tufte, 1999), LegoDB (P. Bohannon, J. Freire, P. Roy, J. Simeon, 2002). Note that this approach depends on the existence of a schema describing the XML data. Furthermore, due to the heterogeneity of XML data, in this approach a simple XML schema/DTD often produce a relational schema with many tables.

Schema-oblivious approach: This method maintains a fixed schema which is used to store XML documents. The basic idea is to capture the tree structure of an XML document. This approach does not require existence of an XML schema/DTD. Also, number of tables is fixed in the relational schema and does not depend on the structural heterogeneity of XML documents. Some examples of schema-oblivious approaches are Edge approach (D. Florescu, D. Kossman, 1999) XRel (M. Yoshikawa, T. Amagasa, T. Shimura, 2001) XParent (M. Yoshikawa, T. Amagasa, T. Shimura, 2001).Schema-oblivious approaches have obvious advantages such as the ability to handle XML schema changes better as there is no need to change the relational schema and a uniform query translation approach. Schema-conscious approaches, on the other hand, have the advantage of more efficient query processing (F. Tian, D. DeWitt, J. Chen, C. Zhang, 2002) Also, no special relational schema needs to be designed for schema-conscious approaches as it can be generated on the fly based on the DTD of the XML document(s).

*The Proposed Method:*

In this section, the structure independent mapping approach (XRecursive) is explained with a sample XML document shown in Figure 1.

*Definition 1:*

XRecursive Structure: XRecursive is a storage structure for storing XML documents where each path is identified by its parent from the root node in a recursive manner.

The proposed method (XRecursive) stores XML document in two different tables; tag_structure table represented the elements associated with its signature and parent element while tag_value table represented the values associated with the elements or type. tag_structure (**tagName, Id, pId**) tag_value (**tagId, Value, Type**)

*XML Document:*

The data structure of XML document is hierarchical, consist of nested structures. The elements are strictly marked by the beginning and ending tags, for empty elements by empty-elements tags. Character data between tags are the content of the elements. It is an instance of XML document contains information about an employee as follows.

```
<? Xml version="1.0" encoding="UTF-8"?>
<Personnel>
<Employee type="permanent">
<Name>Seagull</Name>
<Id>3674</Id>
<Age>34</Age>
</Employee>
<Employee type="contract">
<Name>Robin</Name>
<Id>3675</Id>
<Age>25</Age>
</Employee>
<Employee type="permanent">
<Name>Crow</Name>
<Id>3676</Id>
<Age>28</Age>
</Employee>
</Personnel>
```

**Fig. 1:** XML document.

*Tree Structure:*
   In this section we represented the tree structure of XML document in Fig 1 with XRecursive labeling.

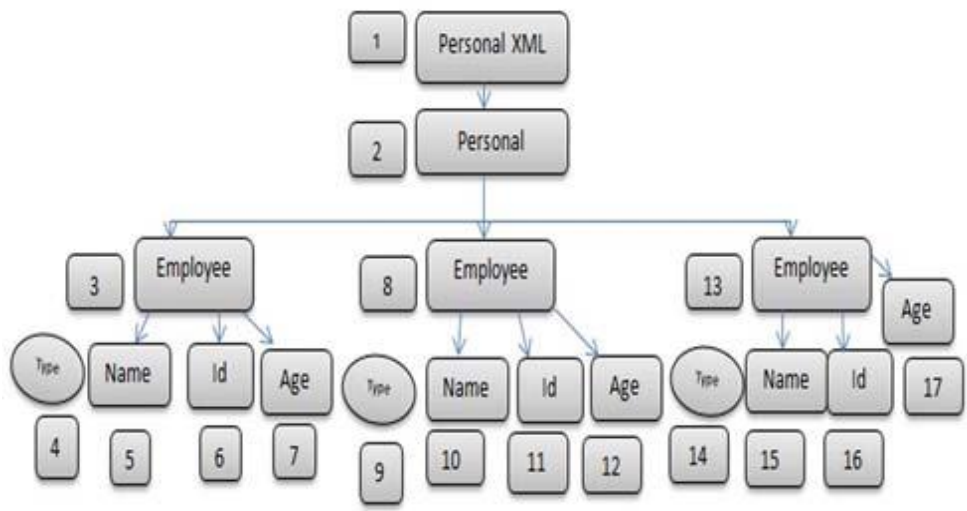

**Fig. 2:** Tree structure of XML document with XRecursive labeling.

*XRecursive Structure:*
   Each and every XML can be describing as a XML tree. In this figure the squares are the elements and the ovals are the attributes of the elements. A generated XML tree has been shown in the figure. Every element or attributes are identified by a signature (number).

*Algorithm:*
   XML document can be stored in relational database, in this paper, MYSQL by use of above two tables. In this paper algorithms are proposed to store XML document into relational database as the following:
*Example 1:*
   In this structure when an element or type associates with its signature it also represents its parent element. We add document name in association with the id to be able to add multiple XML file in the storage. Figure 2 represents the storage of the XML file associated with its signature. For every element there will have a signature associated with it and there will also have a parent's signature associated with it. In table 1: tagName represents the name of the node; id represents the id of the node which is the PK. And finally pId represents the parent id of the node. As document name don't have any parent id so the id of the document name and parent id of the document name is same that has been shown in the table 2.

```
1: Begin
2: let N = { } as empty set, where N represents the list of node of the XML file.
1      Let V = { } as empty set, where V represents the list of the value of the
node.
2      Let String filename = null and int id = 1;

5: filename = read XML document name
6: while xml Document is not null
7: Begin
8: read element name as name
9: read the element type
10: if element type is Element or attribute
11: begin
12: read parent as pName
13: id = id + 1;
14: add name, pName, id to the N
15: End
16: else if element type is #text
17: Begin
18: Read textvalue as value
19: Id = id + 1;
20: Add name, value, id to V
21: end
22: store N into database
23: store V into database
24: End.
```

**Fig. 3:** Algorithm: store XML document to database. **Table 1:** tag_structure.

| tagName | Id | pId |
|---|---|---|
| Personal.xml | 1 | 1 |
| personal | 2 | 1 |
| Employee | 3 | 2 |
| type | 4 | 3 |
| name | 5 | 3 |
| id | 6 | 3 |
| age | 7 | 3 |
| Employee | 8 | 2 |
| type | 9 | 8 |
| name | 10 | 8 |
| id | 11 | 8 |
| age | 12 | 8 |
| Employee | 13 | 2 |
| type | 14 | 13 |
| name | 15 | 13 |
| id | 16 | 13 |
| age | 17 | 13 |

In table 2, we represent the value associated with the elements or type. In XRecursive structure there is no need to store the path value or path structure as it will be determine recursively by its parent id. In Table 1: tagName is the name of the tag, where Id is the parent key. In Table 2: tagId presents the Table 1 id thus tagIdis the foreign key. In Table 2 tagId only represents the elements which contain a value and the value represents on the value column. And the type 'A' denoted to the attribute and 'E' denoted to the element.

**Table 2:** tag_value.

| tagId | Value | Type |
|---|---|---|
| 4 | Permanent | A |
| 5 | Seagua11 | E |
| 6 | 3674 | E |
| 7 | 34 | E |
| 9 | Contract | A |
| 10 | Robin | E |
| 11 | 3675 | E |
| 12 | 25 | E |
| 14 | Permanent | A |
| 15 | Crow | E |
| 16 | 3676 | E |
| 17 | 28 | E |

## REUSLTS AND DISCUSSION

This section discusses the experimental results of storage, parsing and query performance of XML document using XRecursive method. All the experiments were conducted on a Pentium Intel Pentium 4 CPU 3.00 GHz with 1 GB RAM 120GB hard disk. We used Widows XP SP2, java 1.5 SDK and MYSQL 5.1 as the DBMS for storing and retrieving XML document using XRecursive method as structure independent mapping approaches. We have implemented data loader for XRecursive using SAX and DOM parsers. From the results, the proposed method can be use as an efficient way for storing and queering XML data in relational database.

*1-Database Size:*

The database size for XML document in Fig. 1 using XRecursive method is given in show that by this storage method we can reduce not only the size of database requirement of the labeling of node, but also the number of tables.

**Fig. 4:** Database size in KB.

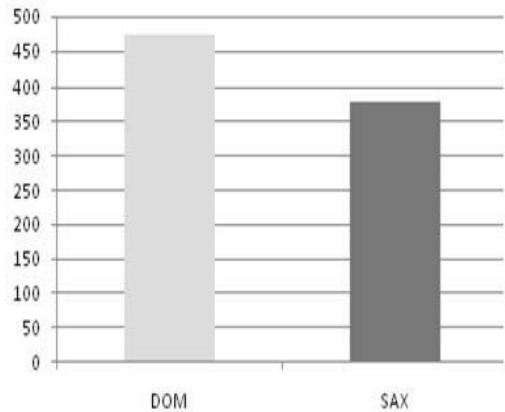

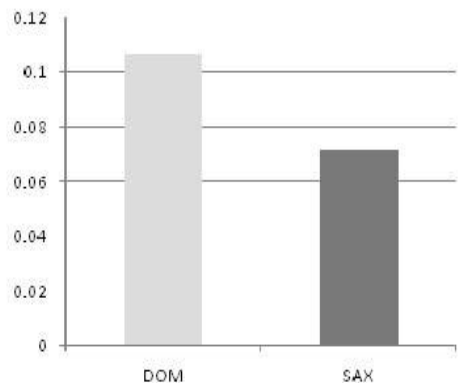

**Fig. 5:** Parsing time in second.

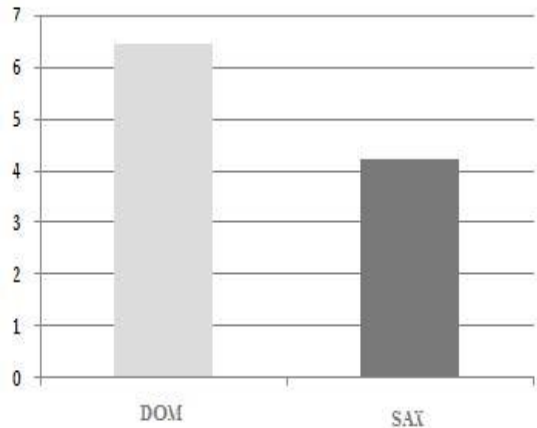

**Fig. 6:** Performance comparison.

*2-Parsing Time:*

The time of parsing the XML document using XRecursive method is faster because it uses Document Object Model (DOM) parsing technique. Using DOM it read the nodes and corresponding child node.

*3-Performance:*

SAX performance is better than DOM and that because DOM uses much more memory Compared to SAX and also because SAX is faster in parsing time than DOM.

*Conclusion:*

XRecursive, a general storage method for XML document using relational database is proposed in this paper. XRecursive adopts the model-mapping method to store XML document in relational database, to decompose the tree structure into nodes and store all information of nodes in relational database according to the node types by recursive way. It can deal with any documents no matter whether it has fixed schema or not. By using this method we can reduce the database size require to store the XML document into relational database. The storing algorithm of XML document into relational database was also given in the paper, and examined the accuracy of it by using the XML document in performance section. Utilizing the actual Xml document evaluated the performance of storing XML document into relational database by using our method.